\documentclass[submission,copyright,creativecommons]{eptcs}
\providecommand{\event}{FIDE 2018} 
\usepackage{breakurl}             
\usepackage{underscore}           
\usepackage{listings}
\usepackage[T1]{fontenc}
\usepackage{beramono}
\usepackage{listings}
\usepackage{graphicx}
\usepackage[table]{xcolor}
\usepackage{tabularx}
\usepackage{lmodern}
\usepackage{amsfonts}
\usepackage[fleqn]{amsmath}
\usepackage{amssymb}
\usepackage{mathtools}
\usepackage{tikz}
\usepackage{tikz-qtree}
\usepackage{bussproofs}
\usepackage{caption}
\usepackage{enumitem}
\usepackage{float}
 
\newtheorem{definition}{Definition}
\definecolor{dkgreen}{rgb}{0,0.6,0}
\definecolor{gray}{rgb}{0.5,0.5,0.5}
\definecolor{mauve}{rgb}{0.58,0,0.82}
\newcolumntype{C}[1]{>{\centering\arraybackslash}p{#1}}

\title{User Support for the Combinator Logic Synthesizer Framework}
\author{Jan Bessai \qquad Anna Vasileva
\institute{Technical University of Dortmund, Germany}
\email{jan.bessai@tu-dortmund.de \qquad anna.vasileva@tu-dortmund.de}
}
\def\titlerunning{User Support for (CL)S}
\def\authorrunning{Jan Bessai and Anna Vasileva}

\begin{document}
\maketitle              
\begin{abstract}
%
Usability is crucial for the adoption of software development technologies.
This is especially true in development stages, where build processes fail, because software is not yet complete or was incompletely modified.
We present early work that aims to improve usability of the Combinatory Logic Synthesizer (CL)S framework, especially in these stages.
(CL)S is a publicly available type-based development tool for the automatic composition of software components from a user-specified repository.
It provides an implementation of a type inhabitation algorithm for Combinatory Logic with intersection types, which is fully integrated into the Scala programming language.
Here, we specifically focus on building a web-based IDE to make potentially incomplete or erroneous input specifications for and decisions of the algorithm understandable for non-experts.
A main aspect of this is providing graphical representations illustrating the step-wise search process of the algorithm.
We also provide a detailed discussion of possible future work to further improve the understandability of these representations.
\end{abstract}
\section{Introduction}

The Combinatory Logic Synthesizer (CL)S Framework provides a publicly available \cite{combinatorsorg} development tool, which is fully integrated into the Scala programming language and can automatically compose software based on types.
Type specifications for (CL)S are based on Combinatory Logic with intersection types \cite{bcl} and automatic software composition is performed by answering the type inhabitation question:
$ \Gamma \vdash ? : \tau $.
In words this question reads: given a set of typed combinators, where each combinator represents a software component, find all applicative terms formed from the combinators in $\Gamma$, which have type $\tau$.
Integration into Scala allows to reuse programming skills and greatly simplifies the specification of combinators.
While user-input is easy to provide and any Scala IDE can be used to program and manipulate combinators, it can be difficult to understand why the algorithm involved in solving the type inhabitation question does or does not produce certain expected solutions.
This is especially true, when (CL)S is used to automatically compose non-trivial software from large component bases, which is its main use-case  \cite{Mixins,FeatureGrammars,CPS,SPLC15,LongWinding}.

According to ISO 9241-11:2018 \cite{iso9241}, the definition of usability is "extent to which a system, product or service can be used by specified users to achieve specified goals with effectiveness, efficiency and satisfaction in a specified context of use". 
In our case, the system, (CL)S, is intended to be used by normal programmers, who are not experts on type theory, and want automatic composition to enhance their efficiency in the context of large collections of software components without being dissatisfied by inexplicable (non-)solutions.
To achieve this aim, a web-based IDE for debugging and improving type specifications is being developed.
It will especially support programmers in development stages, where type specifications are still incomplete, causing the algorithm to find not enough or too many solutions.
An important aspect of this is to visualize the search process performed by the inhabitation algorithm, as well as its -- potentially infinite -- solutions spaces.
The main contributions of this paper are:
to provide an IDE-based view-point on type inhabitation, to investigate how hypergraphs are useful in this context, and to pose a number of interesting research questions.
The paper is organized as follows:
in the remainder of this section we will discuss some related work.
Section~\ref{sec:cls} includes a brief presentation of the (CL)S framework, architecture level aspects of connecting it to an IDE, and more context on the running example of movement combinators.
In Section~\ref{sec:debugger} we discuss the formalism of tree grammars in context of type inhabitation, their relation to hypergraphs and how the IDE presents them to users.
Finally, in Section~\ref{sec:con} we summarize and pose some interesting research questions for future work.

\subsection{Related Work}
\label{sec:rw}
Usability is well-studied and its various aspects \cite{usabilityEngineering} also include the tools for implementing software \cite{programmersAreUsers}. 
Here, we focus on the implementation phase of projects using the (CL)S framework.
A similar effort in the setting of program synthesis and verification has been undertaken within the Leon project \cite{Leon}, which provides a web-based IDE~\cite{LeonIDE}.
Leon uses several SMT-solvers as its back-end. 
It performs verification on the level of individual Scala AST nodes and synthesis is based on invariants.
In contrast, (CL)S has a single back-end algorithm that synthesizes function compositions from type specifications.
The Leon IDE is text-oriented and shows results next to individual lines of code, while graphical result representations are central to our development.
For users preferring text, we annotate each node representing a combinator with source positions indicating its point of origin.
In the area of automatic verification, the Why3 IDE \cite{Why3} is similar to Leon and also textually links the output of various SMT solvers to positions in code.
The RESOLVE programming language \cite{Kabbani}, Coq \cite{WebIDE,jsCoq} and the LEAN theorem prover \cite{Lean} have similar IDEs to support (semi\nobreakdash-)manually solved proof obligations for verified programs.
Resembling our approach and unlike the aforementioned, the Globular proof assistant \cite{Globular} helps to build proofs graphically.
The graphical representation in Globular is based on category theory string diagrams and users manipulate graphs directly, while the hypergraphs in (CL)S are automatically generated.
Globular requires users to be experts in category theory, while (CL)S aims to not require expertise on type-theory.
IDE support is crucial in the complicated process of program verification, and the list of such developments is too long to fully enumerate.
The above examples are selected for being web-based.
In contrast to most web-based IDEs, we do not try to relocate source-code development into the browser, but rather focus on graphically assisting it.
This means, developers can continue using the Scala IDE they are used to.
From the web-based approach we gain platform independence.
At the time of writing, most native client-side user interface libraries require platform specific code or application setup instructions.
Our development works out-of-the-box with a browser and sbt~\cite{sbt}, which is necessary for using (CL)S even without the IDE.

We hope that our work provides useful insights for building IDEs for other settings.
It could especially help with Petri net and hypergraph based synthesis~\cite{petriHG}.
To our knowledge, there is no other tool to debug intersection type specifications~\cite{CoppoDezani}, which are an important area of research~\cite{ITRS,HindleyIntro} with multiple recent applications to synthesis~\cite{BoundedDimension,NonIdempotent,ExampleBasedSynthesis} beyond Combinatory Logic~\cite{bcl}.

\section{An Overview of (CL)S Scala Framework}
\label{sec:cls}
\includegraphics[width=\textwidth]{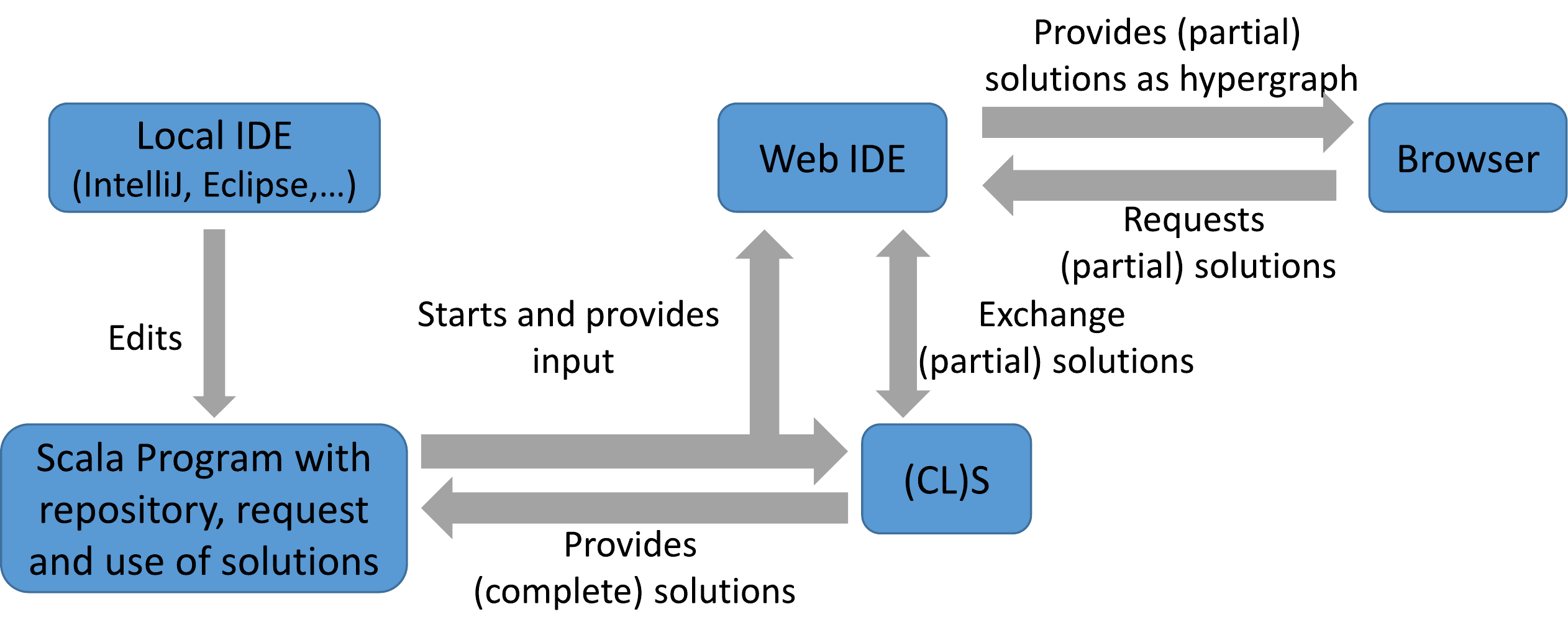}
The illustration above provides an overview of the data flow when using and debugging specifications in (CL)S.
Programmers specify and implement combinators in Scala using any local IDE.
Their programs also include synthesis targets and put results of the algorithm to further use, e.g. by interpreting combinator applications as function calls.
The framework itself can be used as a library from any Scala program.
If the additional debugging capabilities of the web-based IDE are desired, it is started by instantiating a controller for the Play web framework \cite{Play}.
Starting the application will instantiate a web server, which hosts a web site where users can access debugging information, e.g. in form of hypergraph-based visualizations of the search process.
Behind the scenes, the debugger communicates with the algorithm implementation provided by (CL)S, which has appropriate interfaces to observe its internal state.

There are four basic rules to control the process of type inhabitation \cite{bcl}.
The first rule (if $c : \tau \in \Gamma$ then $\Gamma \vdash c : S(\tau)$) allows to use any combinator $c$ present in repository $\Gamma$ with type $\tau$ and to assume that it has type $S(\tau)$, where $S$ is a substitution mapping type variables in $\tau$ to types without variables.
This way, e.g. the identity function can have type $\alpha \to \alpha$ that can turn into $Int \to Int$ or $String \to String,$ depending on the choice of substitution.
In order to make type inhabitation decidable, substitutions are drawn from a finite space (instead of guessed), which is part of the input specification.
The second rule (if $\Gamma \vdash M : \sigma \to \tau$ and $\Gamma \vdash N : \sigma$ then $\Gamma \vdash M N : \tau$) allows to apply combinators with function types to appropriately typed arguments.
Moreover, we have the intersection introduction rule ($\Gamma \vdash M : \sigma$ and $\Gamma \vdash M : \tau$ implies $\Gamma \vdash M : \sigma \cap \tau$)  to type the same term with two types, if both types can be derived.
Lastly, the type system supports subtyping ($\Gamma \vdash M : \sigma$ and $\sigma \leq \tau$ implies $\Gamma \vdash M : \tau$).
When answering an inhabitation question, (CL)S will search for combinators and their arguments, such that synthesized applicative terms are well-typed.
Input specifications are the combinator repository, allowed substitutions for type variables (if there are any), a subtype relation on type constants to define $\leq$, and the requested type.

We consider labyrinths as an example and try to find all paths from an entrance to a goal. 
The following labyrinth has start position $(0, 2)$ and goal position $(1, 0)$:
\begin{center}
\begin{tabular}{cc}
\begin{tabular}{r|C{0.4cm}|C{0.4cm}|C{0.4cm}|}
  & 0                 & 1                 & 2 \\\hline
0 & \cellcolor{black} & $\bigstar$        & \cellcolor{black} \\\hline	
1 &                   &                   &  \\\hline
2 &	$\bullet$         &	\cellcolor{black} &\\\hline
3 &                   &                   &\\\hline
\end{tabular}&
\hspace{-2em}$\begin{aligned}
    \Gamma_{ex} = \{   
           left  :\;& (Pos(1, 1) \to Pos(0, 1)) \cap Pos(2, 1) \to Pos(1, 1))\ \cap \\
                    & (Pos(1, 3) \to Pos(0,3)) \cap (Pos(2, 3) \to Pos(1, 3)),\\
           right :\;& (Pos(0, 1) \to Pos(1, 1)) \cap (Pos(1, 1) \to Pos(2, 1))\ \cap \\
                    &(Pos(0, 3) \to Pos(1,3)) \cap (Pos(1, 3) \to Pos(2, 3)),\\
            up :\;& (Pos(0, 3) \to Pos(0, 2)) \cap (Pos(2, 3) \to Pos(2, 2))\ \cap \\
            	&(Pos(1, 1) \to Pos(1, 0)) \cap (Pos(0, 2) \to Pos(0, 1))\ \cap \\
            	&(Pos(2, 2) \to Pos(2, 1)), \\
           down  :\;& (Pos(1, 0) \to Pos(1, 1)) \cap (Pos(0, 1) \to Pos(0,2))\ \cap\\
           		&(Pos(2, 1) \to Pos(2, 2)) \cap (Pos(0, 2) \to Pos(0,3))\ \cap \\
           		&(Pos(2, 2) \to Pos(2, 3)),\;\; start :\; Pos(0, 2)\quad\}                 
\end{aligned}$
\end{tabular}
\end{center}

\noindent There are multiple possible paths to reach the goal position. 
Repository $\Gamma_{ex}$ shows the labyrinth in mathematical notation.   
Each entry in $\Gamma_{ex}$ consists of a combinator name and its type description. 
The repository represents the start position and all possible one-step moves as typed combinators.
Types indicate column and row positions in the labyrinth.
Combinators for going up, down, left or right are functions from a start to a destination position.
Intersection types allow movement combinators to have multiple types at once, e.g. combinator $up$ can be used to go from $Pos(1, 1)$ to $Pos(1, 0)$ and $Pos(2, 2)$ to $Pos(2, 1)$.
To make the example more readable, we avoid variables, which would have allowed for specifications like $up : Pos(\alpha, PlusOne(\beta)) \to Pos(\alpha, \beta)$.
We get all possible paths through the labyrinth by asking for the goal position, e.g. $\Gamma_{ex} \vdash ? : Pos(1, 0)$.
The algorithm computes all solutions in form of tree grammars which will be shown as hypergraphs.

\section{IDE for the (CL)S Framework}
\label{sec:debugger}
The (CL)S framework recursively grows the set of production rules of a tree grammar to describe solutions. 
Tree grammars are well-known from literature \cite{tata,treeG} and we consider the generalized case of regular tree grammars without restrictions on the arity of terminal symbols.
Formally we have:
\begin{definition} A tree grammar G is a 4-tuple $(S, N, \mathcal{F}, R)$ with a start symbol $S \in N$, a set $N$ of \textit{nonterminals}, a set $\mathcal{F}$ of \textit{terminal symbols}, and a set $R$ of \textit{productions rules} of form $\alpha \mapsto f(\beta_1, \beta_2, \dots \beta_n)$, where $n \geq 0$, $\alpha, \beta_1, \beta_2, \dots, \beta_n \in N$ are nonterminal and $f \in \mathcal{F}$ is terminal.
For a given tree grammar $G = (S, N, \mathcal{F}, R)$ and nonterminal $\alpha \in N$, $\mathcal{L}_{\alpha}(G)$ is the least set closed under the rule:
$ \begin{aligned}[t]
    &\text{ if } \alpha \mapsto f(\beta_1, \beta_2, \dots, \beta_n) \in R \text{ and for all } 1 \leq k \leq n: t_k \in \mathcal{L}_{\beta_k}(G) \text{ then }\\
    &f(t_1, t_2, \dots, t_n) \in \mathcal{L}_{\alpha}(G).
\end{aligned}$\\
We define $\mathcal{L}(G) = \mathcal{L}_{S}(G)$ to be the language of grammar G.
\end{definition}
Solutions in (CL)S are terms $M$, which are well-typed for a requested type $\tau$ relative to the type assumptions explained above.
Given $\Gamma$ and $\tau$, (CL)S constructs a tree grammar $G = (\tau, N, \mathcal{F}, R)$ such that $\tau \in N$ and for all $\sigma \in N$ we have $M \in \mathcal{L}_{\sigma}(G)$ if and only if $\Gamma \vdash M : \sigma$.
In other words, we get a tree grammar where right hand sides of rules start with a combinator symbol followed by the types of arguments required to obtain the type on the left hand side of the rule by applying the combinator.
The start symbol is the user requested target type.

\noindent Let us consider the following example labyrinth to illustrate the search process:
\begin{center}
\begin{tabular}{cc}
\begin{tabular}{r|C{0.4cm}|C{0.4cm}|C{0.4cm}|C{0.4cm}|C{0.4cm}|}
\multicolumn{5}{c}{}\\
  & 0        & 1                 & 2     & 3                 & 4\\\hline
0 &$\bullet$ & \cellcolor{black} & $\bigstar_2$ &                   & \cellcolor{black}\\\hline
1 &$\bigstar_1$     & \cellcolor{black} &       & \cellcolor{black} & $\bigstar_3$ \\\hline
\end{tabular}&
$\begin{aligned}[c]
&\;\;\begin{aligned}[t]
      \Gamma = \{
      up    :\;& (Pos(0, 1) \to Pos(0, 0)) \cap (Pos(2, 1) \to Pos(2,0)),\\
      down  :\;& (Pos(0, 0) \to Pos(0, 1)) \cap (Pos(2, 0) \to Pos(2,1)),\\
      left  :\;& Pos(3, 0) \to Pos(2, 0),\\
      right :\;& Pos(2, 0) \to Pos(3, 0),\; start :\; Pos(0, 0) \quad\}
 \end{aligned}
\end{aligned}$
\end{tabular}
\end{center}
For goal position $\bigstar_1$, we ask $\Gamma \vdash\;? : Pos(0, 1)$ and combinator $down$ can be used with argument $Pos(0, 0)$.
The first computed tree grammar entry will be $Pos(0, 1) \mapsto down(Pos(0, \omega) \cap Pos(\omega, 0))$. 
For internal implementation reasons of the search procedure, (CL)S chooses the nonterminal for the argument of down to represent a type, which is subtype equal to $Pos(0, 0)$ (in Combinatory Logic with intersection types we have: $Pos(0, \omega) \cap Pos(\omega, 0) \leq Pos(0, 0) \leq Pos(0, \omega) \cap Pos(\omega, 0)$, because type constructors like $Pos$ are co-variant in their arguments, distribute over intersection and $\omega$ is the universal supertype of every other type).
Type $Pos(0, \omega) \cap Pos(\omega, 0)$ will become the next target, for which two rules are computed:
$Pos(0, \omega) \cap Pos(\omega, 0) \mapsto start()$, which is obvious, and $Pos(0, \omega) \cap Pos(\omega, 0) \mapsto up(Pos(0, 1))$, which is perhaps surprising.
The computed tree grammar is not only sound (its words are well-typed terms), but also complete (all requested well typed terms are words).
Hence, the cyclic second rule causes terms like $down(up(down(start))), down(up(down(up(down(start))))), \dots$ to be derivable.
Inhabitation stops, because a rule for the argument of $up$ can be found in the already computed grammar ($Pos(0, 1)$ is the left hand side of the first rule) and no further recursive targets exist.
For goal position $\bigstar_2$ the algorithm computes rules $R = 
  \{ Pos(2, 0) \mapsto up(Pos(2, \omega) \cap Pos(\omega, 1)),$ $
     Pos(2, 0) \mapsto left(Pos(3, \omega) \cap Pos(\omega, 0)),$ $
     Pos(2, \omega) \cap Pos(\omega, 1) \mapsto  down(Pos(2, 0)),$ $
     Pos(3, \omega) \cap Pos(\omega, 0) \mapsto right(Pos(2, 0)) \}.$
This time rules are cyclic and unproductive, no word can be derived for the start symbol $Pos(2, 0)$.
In the implementation, all unproductive rules are pruned from the grammar and one motivation for having a debugger is to inform users about that process.
Another motivation arises when considering goal $\bigstar_3$, for which no combinator exits.
The tree grammar will be empty and users would have to check the entire repository to find out why.

We introduce the \textit{hypergraphs} \cite{hg} as a graphical representation of tree grammars.
\begin{definition}
A directed labeled hypergraph H over an alphabet $\mathcal{F}$ is a 5-tuple H = (V, E, nod$_E$, nod$_V$, lab) where V is a finite set of nodes, E is a finite set of hyperedges, incidence is specified by a function nod$_E: E \to V^{*}$ and a relation nod$_V \subseteq V \times E$, and labels are given by a function lab$: E \to \mathcal{F}$.
\end{definition}
Every edge in a hypergraph has outgoing connections described by nod$_E$ and incoming connections described by nod$_V$.
Outgoing connections are finite vectors of nodes.  
This is different from normal graphs, where edges only connect two nodes. 
All tree grammar nonterminals are represented by nodes ($V = N$). 
For each production $\alpha \mapsto f(\beta_1, \beta_2, \dots \beta_n)$ we add an edge $e$ with nod$_V(\alpha) = e$ and $(e, (\beta_1, \beta_2, \dots \beta_n)) \in$ nod$_E$ where lab$(e) = f$.

Figures \ref{fig:x1} and \ref{fig:x2} provide an overview of the hypergraphs for the tree grammars for the prior labyrinth example.
The hypergraph for position $\bigstar_1 = Pos(0, 1)$ in Fig.~\ref{fig:x1} contains nodes for every nonterminal (type) and combinator usages are modelled by edges.
Outgoing edge connections are numbered to indicated argument positions. 
We draw nodes and edges using boxes and circles. 
The cycle between combinators $up$ and $down$ is immediately visible in Fig.~\ref{fig:x1}.
We can escape the cycle using edge $start$, which has no outgoing connections and thereby models use of a combinator not depending on additional recursively synthesized targets. 
The labels collected for all ways through the hypergraph from $start$ to $Pos(0,1)$ are words of the grammar and valid movement instructions through the labyrinth.   
For $\Gamma \vdash ? : Pos(2,0)$, the IDE shows a message that there is no solution.
In this case, the user can comprehend the problem by means of the visualization provided by the debugger mode.
Figure~\ref{fig:x2} shows the cyclic rule obtained in the last step. 
By comparing the graphs, we see that this hypergraph is different form the first one (Fig.~\ref{fig:x1}): there is no edge without outgoing connections to break cycles.
In the visualization, all edges involved in unbreakable cycles are marked in red.

\begin{figure}[H]
\begin{center}
\includegraphics[width=6cm]{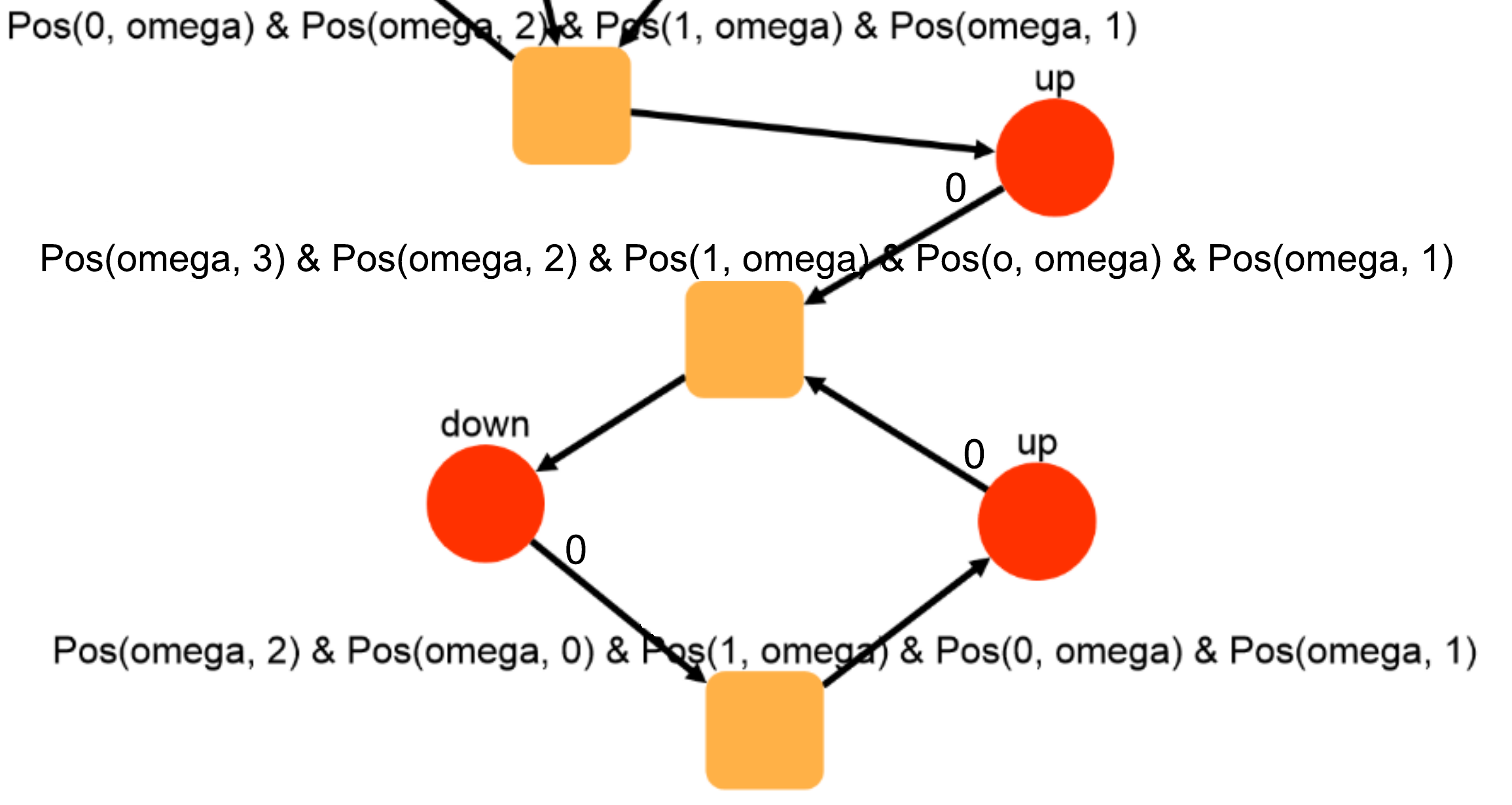}
\caption{Unproductive cycle in solution construction}
\label{fig:cycle}
\end{center}
\end{figure}

\begin{minipage}[t]{0.44\textwidth}
\includegraphics[width=\textwidth]{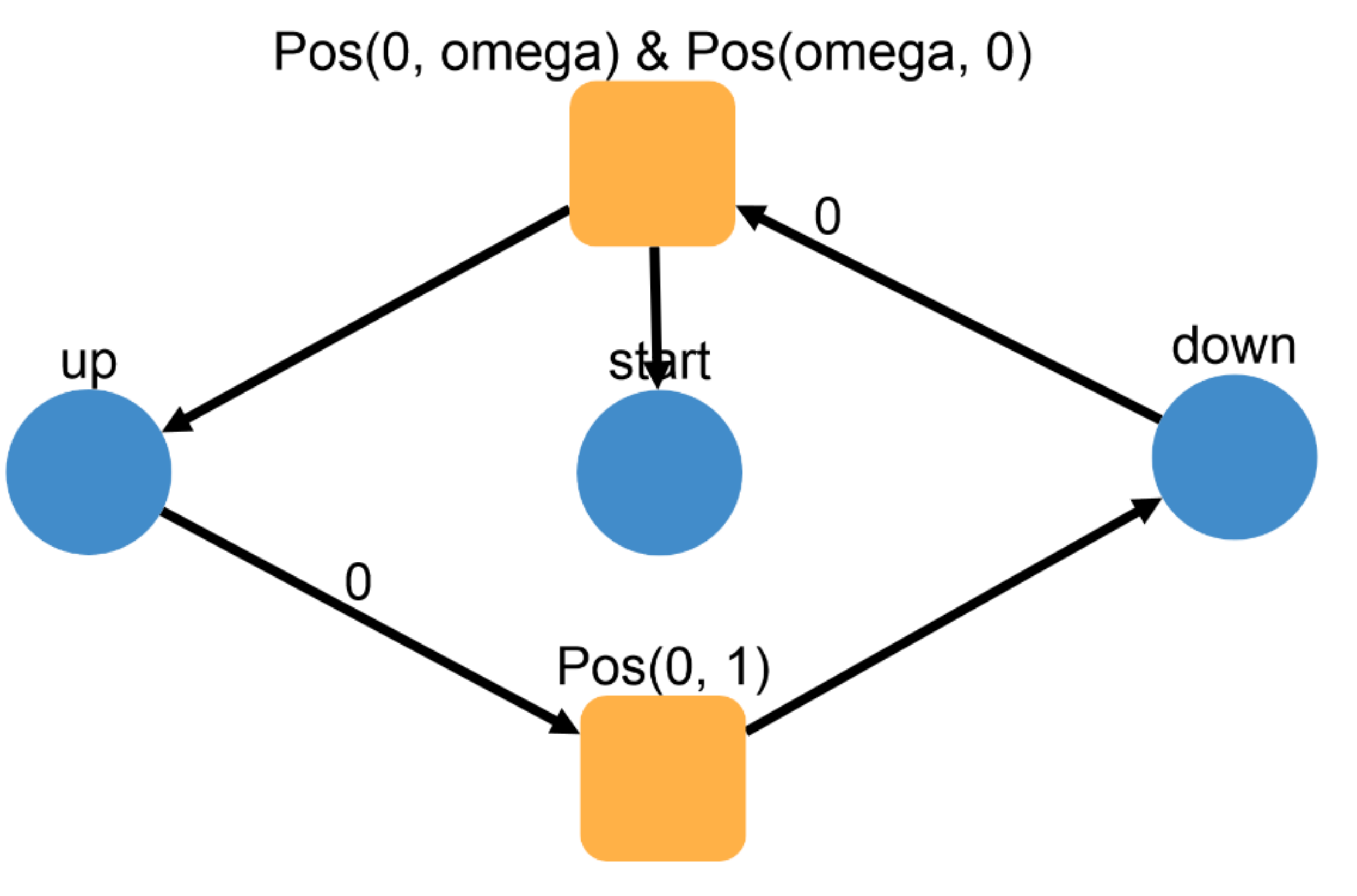}
\captionof{figure}{$Pos(0,1)$}
\label{fig:x1}
\end{minipage}
\begin{minipage}[t]{0.44\textwidth}
\includegraphics[width=\textwidth]{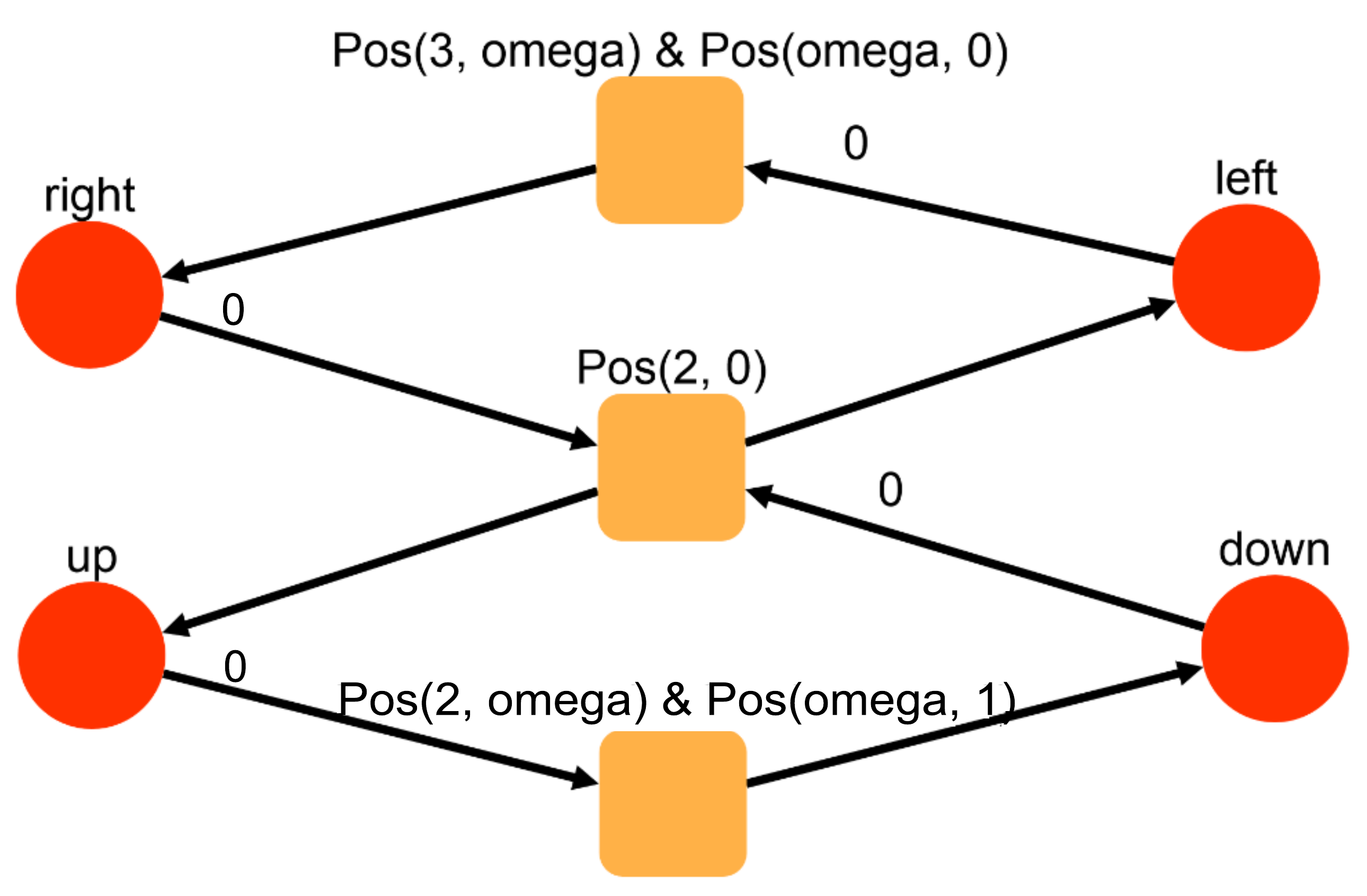}
\captionof{figure}{$Pos(2,0)$}
\label{fig:x2}
\end{minipage}
\begin{figure}[H]
\begin{center}
\frame{\includegraphics[width=\textwidth]{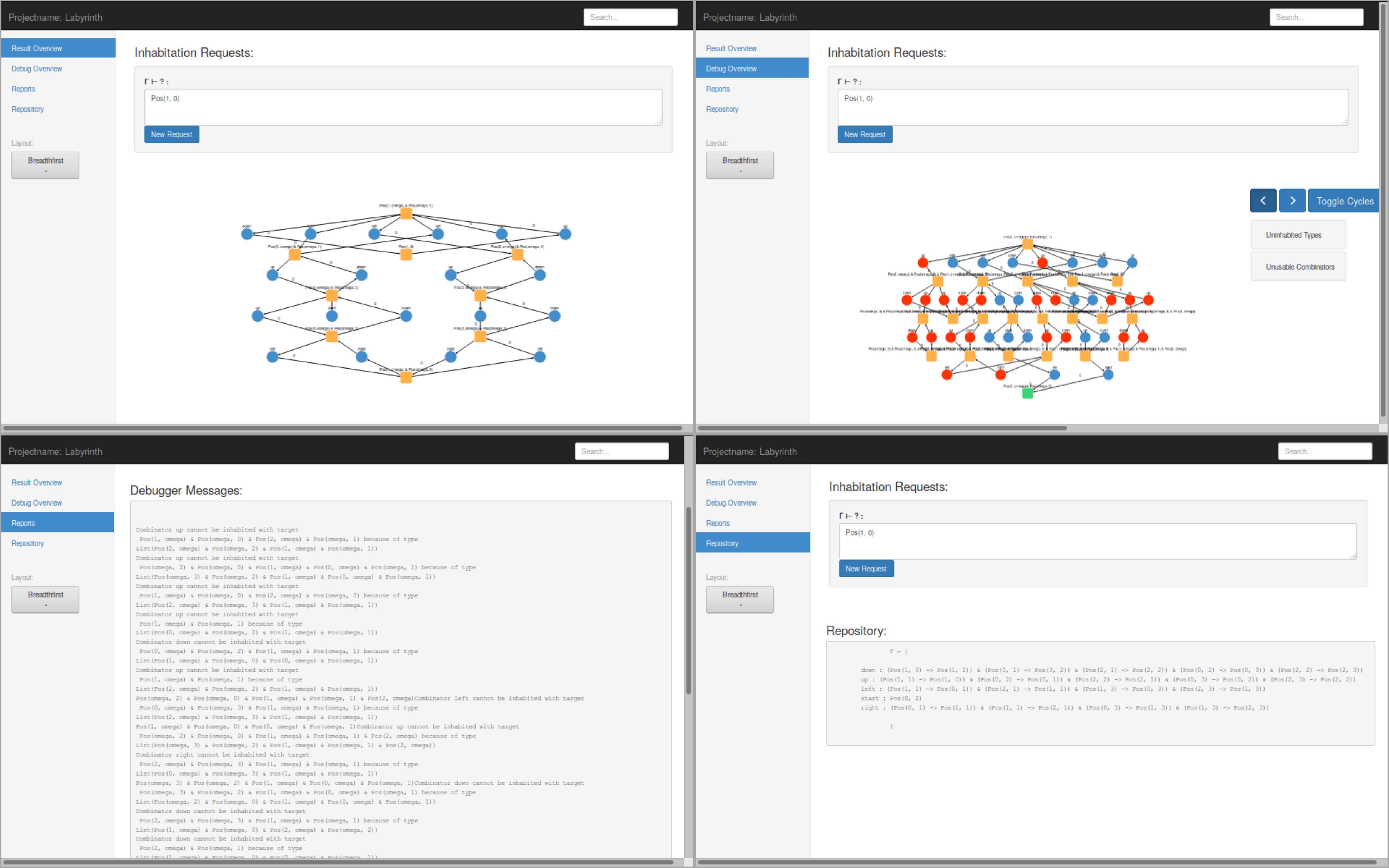}}
\captionof{figure}{Debug Overview}
\label{fig:debugOverview}
\end{center}
\end{figure}
The web-based IDE provides four different perspectives (Fig.~\ref{fig:debugOverview}). 
The \textit{Result Overview} presents the solutions of the inhabitation problem in form of a hypergraph. 
In order to inform about unsuccessful inhabitation, a message is shown instead of the empty hypergraph. 
The \textit{Debug Overview} (Fig.~\ref{fig:debugOverview}, top right) and \textit{Reports} (Fig.~\ref{fig:debugOverview}, bottom right) perspectives provide detailed graphical and textual information about the inhabitation process, which otherwise can be unexpected and incomprehensible. 
Users can access their specification in the \textit{Repository} (bottom right) perspective.     
The \textit{Result Overview} perspective also provides the possibility to make a new request by means of the browser-based IDE. 
This enables fast user interactive experiments with types different (e.g. more specific or generic) from the programmatic request stated in Scala. 
For the visualization of the hypergraph construction, we use the open-source JavaScript library Cytoscape \cite{cy}, which facilitates the fast and interactive representation of hypergraphs exchanged in a simple JSON format.
User interactiveness allows to zoom into a graph as well as to move the nodes and edges.
Layout choices are supported by eight different automatic layout algorithms~\cite{cy}.
We use the Bootstrap HTML components \cite{bootstrap} to gain platform and browser independence.
In the \textit{Debugger Overview} perspective, users can see the generation of solutions in a step-wise process. 
Figure~\ref{fig:debugOverview} (top right) shows a step of the construction process of the solution for the labyrinth example $\Gamma_{ex}$ presented in Section~\ref{sec:cls}.
In the current step, we see that there is also an unproductive cycle. 
Figure~\ref{fig:cycle} shows a zoomed-in and manually re-layouted part of the graph with a cycle in this step. 
Because of the intersection type specific rules, combinator \textit{up} can be used with type $(Pos(\omega, 3) \cap Pos(\omega, 2)\cap Pos(1, \omega) \cap Pos(0, \omega) \cap Pos(\omega, 1)) \to (Pos(0, \omega) \cap Pos(\omega, 2)\cap Pos(1, \omega) \cap Pos(\omega, 1))$.
This is surprising for non-experts.
However, it does not lead to invalid solutions, because all possible inhabitants for the argument type of \textit{left} are generated from an unproductive cycle.
We include a button to toggle all unproductive cycles providing a clean unsurprising view.
There are no combinators for goal $\bigstar_3$,  therefore the associated hypergraph is uninformative. 
The graph in the \textit{Debug Overview} perspective contains only type $Pos(4,1)$ as a green node, with color green indicating yet to-do recursive targets. 
Since there exists no suitable combinator in $\Gamma$, the graph for the next step is empty.
Moreover, users can find information in the \textit{Reports} perspective.
It includes textual information about each uninhabited type encountered during the search process.
Reasons for non-inhabitation (unproductive cycles, no usable combinators) are distinguished.
The \textit{Repository} perspective gives users an overview of their repository specifications.
These are not always entered manually, but can also be programmatically constructed from a problem domain.
In case of the labyrinth examples, it is easy to write a Scala program to create $\Gamma$ from a two dimensional boolean array.



\section{Conclusion and Future Work}
\label{sec:con}
The presented IDE for (CL)S is geared toward debugging incomplete or erroneous input type specifications.
We provided an overview of the framework and IDE together with an easy to understand application example.
The example is then used to illustrate the connection between tree grammars generated by (CL)S and hypergraphs, which are used for visualization in the debugger.
We have shown how intermediate synthesis steps can be visualized and unproductive cycles can be seen in the hypergraph.
This step-wise visual construction of the hypergraphs may help non-experts understand the decisions of the algorithm.
There are numerous areas of future work: 
\begin{description}[leftmargin=0em]
\item[Evaluation]
While the debugger helped us to understand and debug our own examples, more evaluation is needed to see if it really helps non-experts.
A possible evaluation scenario would include student groups and measure their effectiveness in solving a given task~\cite{CodeQuality}.
It would also be important to see which further features would be needed to make (CL)S scale to development teams.
This is especially interesting, when developers try to understand type-specifications devised by others.
There is little to no work on studying this subject in the area of software synthesis in general.
\item[Performance] While writing this paper we were able to greatly improve the performance of (CL)S, because the hypergraphs constructed for the labyrinth example revealed the generation of redundant recursive inhabitation targets.
Due to the computational complexity of type inhabitation in Combinatory Logic with intersection types, which is above EXPTIME \cite{bcl}, we expect that there always will be scenarios where users have to wait for results.
Some further optimizations have been present in an earlier F\# based version of (CL)S, and just need to be ported.
For drawing and layouting of hypergraphs, Cytoscape was fast enough for user interactive operation, but we expect limitations when solutions get too big.
In this case, partial collapsing of hypergraphs might help.
\item[Input specification quality] Input specifications can include badly designed combinators.
Indications of bad design would be unusable combinators, unnecessarily generic or overly specific types, or overly long parameter lists leading to implementation code smells.
We plan to improve our IDE to provide user feedback by statically analyzing the repository.
This kind of analysis will potentially give further insights on intersection types, relying on practical applications of techniques such as intersection type matching or unification \cite{IntersectionTypeUnification}.
It might lead to a formal understanding of what it means to be a component suitable for synthesis and reveal a potential connection to clean code in regular programming.
\item[Algorithmic Artifacts] We have seen that the search procedure will introduce hard to understand type artifacts, representing type $Pos(0, 0)$ as $Pos(0, \omega) \cap Pos(\omega, 0)$.
The cycle shown in Fig.\ref{fig:cycle} is such an artifact, because its types were created by the intersection introduction rule, which can have surprising consequences in the case of large intersections.
A detailed case-to-case analysis of intersection types in practical experiments will be necessary to find good countermeasures to the aforementioned problems.
Most other synthesis techniques rely on some form of intermediate representation and clearly more research is needed on how to make the connection of these representations to the initially specified problem obvious to non-experts.
\end{description}
In addition to these future work topics the structure of Petri nets can be represented as directed hypergraphs \cite{PetriNetsHyperGraphs} and thus the iterative construction in \cite{petriHG} could be adapted to be shown in our IDE.
In reverse, recently developed web-based tools for debugging and benchmarking Petri nets \cite{PetriNetDebugging} have the potential to give useful input for our future work.
\bibliographystyle{splncs04}
\bibliography{bibliography}
\end{document}